\documentclass[pra, aps, 10pt, twocolumn, floatfix,superscriptaddress]{revtex4-1}

\usepackage{graphicx}
\usepackage{amssymb}
\usepackage{amsmath}
\usepackage{color}
\usepackage{changes}
\usepackage[normalem]{ulem}

\begin{document}

\title{Excitation and control of quantum well nanostructures by unipolar half-cycle attosecond pulses}

\author{Rostislav Arkhipov}
\affiliation{St. Petersburg State University, Universitetskaya nab. 7/9, St. Petersburg 199034, Russia}
\affiliation{Ioffe Institute, Polytekhnicheskaya 26, St. Petersburg, 194021, Russia}

\author{Pavel Belov}
\affiliation{Universit\"{a}t Rostock
Albert-Einstein-Strasse 23, 18059 Rostock, Germany}
\affiliation{St. Petersburg State University, Universitetskaya nab. 7/9, St. Petersburg 199034, Russia}

\author{Anton Pakhomov}
\affiliation{St. Petersburg State University, Universitetskaya nab. 7/9, St. Petersburg 199034, Russia}

\author{Mikhail Arkhipov}
\affiliation{St. Petersburg State University, Universitetskaya nab. 7/9, St. Petersburg 199034, Russia}

\author{Nikolay Rosanov}
\affiliation{St. Petersburg State University, Universitetskaya nab. 7/9, St. Petersburg 199034, Russia}
\affiliation{Ioffe Institute, Polytekhnicheskaya 26, St. Petersburg, 194021, Russia}

\begin{abstract}

Unipolar and quasi-unipolar half-cycle pulses having nonzero electric pulse area are a limit of pulse shortening in a given spectral range. In spite of the fact that existence of such pulses was considered by Jackson (1962), V.L. Ginzburg (1960-s), Bullough and Ahmad (1971) as well as Bessonov (1981), the possibility of their existence and propagation in space remained questionable for many years. Only within the past decades both the possibility of unipolar pulse existence and their propagation dynamics were shown and analyzed in detail theoretically and experimentally. So far such pulses became a subject of active research due to their potential in the ultra-fast optics and study of new regimes of light-matter interactions with subcycle resolution. Here, we show the possibility of the effective ultrafast control of the level populations in quantum well nanostructures by the half-cycle unipolar attosecond light pulses in comparison to the single-cycle ones. It is shown that the population dynamics can be determined by the electric pulse area divided to its characteristic "scale" determined by the quantum well width. The selective excitation of quantum states and the feasibility of the population inversion by the subcycle unipolar pulses is demonstrated.
\end{abstract}

\maketitle

\section{Introduction}

Recently, the ultra-short pulses with the attosecond durations have become available~\cite{Krausz, Calegari, Hassan, biegert2021attosecond, midorikawa2022progress}. Due to the lengths comparable with characteristic periods of the electron state eigenfrequencies
in atoms, molecules and solids such pulses constitute a remarkable tool of the ultrafast control of wavepacket properties~\cite{Calegari, Ramasesha, Hassan, nisoli2017attosecond,yang2022comparative,freudenstein2022attosecond}. Recent experimental results showed the possibility to control the absorption spectral lineshape in molecules~\cite{peng2022coherent} and to realize spatial time imaging of the electronic coherence angstrom-scale spatial and subfemtosecond temporal resolutions~\cite{garg2021real}. Furthermore, an interaction of the few-cycle pulses with nanostructures (e.g. nanospheres, quantum wells, quantum dots) is becoming a relevant topic of modern optics now since they used to control the ultrafast nonlinear processes at the nanoscale, see reviews~\cite{ciappina2017attosecond,seiffert2022strong,ciappina2023multiphoton}.

Conventional femtosecond and attosecond pulses, generated from the mode-locked lasers and high-order harmonics generation setup, contain several cycles of oscillations within the pulse duration ~\cite{Krausz, Calegari, Hassan, biegert2021attosecond, midorikawa2022progress}. They are bipolar: the electric field strength vector changes its direction to opposite several times during the pulse duration. Then, the electric pulse area~\cite{Rosanov_Usp}, defined as an integral of the electric field strength $E(t)$ with respect of time $t$ at a given point of space,
\begin{equation}
\label{introS}
 \vec S_{E} = \int_{t=-\infty }^{+\infty} \vec E(t) dt,
\end{equation}
is always close to zero and does not make any sense. That is why this quantity has not been considered so far.

A limit of the pulse shortening in the corresponding spectral range is a single-cycle pulse containing one period of the electromagnetic oscillation - two half waves of the opposite polarity. Further shortening leads to a so-called unipolar half-cycle pulse (UP) possessing a single-sign half-wave of the electric field ~\cite{Arkhipov_JETP,Rosanov_review,arkhipov2020unipolar}. In such a case, the electric area~(\ref{introS}) of the unipolar pulse can be not equal to zero.

The existence of unipolar pulses was questionable for many years. Although the propagation of finite-energy unipolar pulses in an unbounded free space without any charges and current sources is impossible, they can propagate with the speed of light in vacuum inside hollow coaxial waveguides possessing no cutoff frequencies \cite{Rosanov2019}. Earlier existence of such pulses with nonzero electric area generated by moving relativistic charges  was considered by Jackson \cite{jackson1962classical} (Ch. 11) and Bessonov  \cite{Bessonov}. Existence of unipolar solitons travelling in a medium was shown by Bullough and Ahmad \cite{Bullough}. Rosanov derived directly from the Maxwell equations for the first time the electric field area conservation law in 1D problems $\frac{d}{dz} \vec{S}_E=0$  \cite{rosanov2009area,Kozlov}. In 3D-geometry this rule takes the form $\text{rot} \ \vec{S_E}=0$, see \cite{Rosanov_Usp} and references therein. Consequences from this rule and the propagation dynamics of a half-cycle pulse in a resonant medium were discussed in \cite{arkhipov2021coherent}. Propagation dynamics of a unipolar THz pulse in the 3D free space is studied in \cite{sychugin2022propagation}.

Some examples of pulses with non-zero electric area (\ref{introS}) are provided both in some theoretical \cite{Kozlov,rosanov2020formation,rosanov2020electric, plachenov, Rosanov_UFN_2023} and experimental studies \cite{Naumenko, naumenko2020unipolar, arkhipov2021experimental}, see also the reviews ~\cite{Rosanov_review, arkhipov2020unipolar}. A train of half-cycle unipolar pulses can be obtained via synchrotron radiation as shown by V.L. Ginzburg et al. \cite{ginzburg2013origin,ginzburg1966cosmic,ginzburg1968synchrotron,ginzburg2013theoretical}. Besides the formation of half-cycle attosecond pulse train is possible when two-color laser pulses interact with atoms \cite{persson2006generation}. A historical overview of the earlier results on such pulses can be found in the review \cite{sazonov2023nonlinear} and references therein.

An interest to obtaining of unipolar and quasi-unipolar pulses and their interaction with matter has increased rapidly within the last several years \cite{arkhipov2020unipolar,arkhipov2022half,arkhipov2023unipolar}. Unipolar and quasi-unipolar pulses in THz range can be obtained via a nonlinear amplification of the initial bipolar THz pulse in the non-equilibrium plasma channel \cite{bogatskaya0,bogatskaya1,bogatskaya2,bogatskaya2023possibility}, via transition radiation of a laser-produced relativistic electron bunch~\cite{Kuratov} or via coherent control of medium polarization by femtosecond pulses~\cite{Pakhomov,Pakhomov_SciRep, Pakhomov_PRA_2022}. Experimentally, the unipolar THz pulse can be generated in the near field from ultra-fast laser plasma in solid targets \cite{Gao}, in filaments \cite{Arkhipovunipolar}, in the semiconductor media in the form of  a precursor~\cite{Tsarev}. Besides experimental evidence of unipolar pulse formation via diffraction radiation of relativistic electron beam was given in \cite{Naumenko,naumenko2020unipolar} Also, solitonic solutions of the nonlinear optics equations in the form of unipolar pulses exist~\cite{Bullough,sazonov2021unipolar,sazonov2022soliton}. Besides, the formation of a unipolar half-cycle pulse  from an initially bipolar one in a nonlinear medium was theoretically examined in \cite{Kalosha,Kozlov,Song}. The half-cycle attosecond pulses having a strong half-wave of one polarity in the optical range can be obtained via the fast acceleration of electrons in a foil target~\cite{Wu, Xu, Eliasson} or Fourier synthesis of broadband pumping \cite{Hassan}. More detailed review on recent advances in the physics of unipolar pulses can be found in recent reviews~\cite{Arkhipov_2022_LPL,arkhipov2023unipolar}.

So far, most studies in optics have dealt with the multi-cycle excitation of atomic systems and nanostructures~\cite{Calegari, Ramasesha, Hassan, nisoli2017attosecond,yang2022comparative, ciappina2017attosecond, seiffert2022strong}, for example, using the monochromatic resonant excitation.   In particular resonant excitation of quantum systems by multi-cycle field is used \cite{Allen}. Interaction of unipolar field with atomic systems is nonresonant and significantly differs from commonly used resonant multi-cycle light-matter interaction. Indeed, a remarkable feature of unipolar half-cycle pulses is related to the presence of a monopolar burst of the electric field, which leads to their unidirectionality \cite{Song}. Usual  bipolar multi-cycle pulses act on the electron in forward and backward direction and the total displacement after the pulse is not significant. 
Instead, a unipolar half-cycle pulse transfers the kinetic momentum to electron along one direction.
As the result, as compared to few-cycle pulses unipolar ones can effectively transfer kinetic momentum (short "kick") to free or bound charges in atoms. This momentum transfer is especially efficient, if the pulse duration $\tau_p$ is shorter than the characteristic time $T_g = 2\pi\hbar/E_1$ associated with the energy of the ground state $E_{1}$ (the classical electron orbital period). 
The described property of unipolar pulses makes them the perfect tool for efficient non-resonant excitation and ultrafast control of wavepacket dynamics in media ~\cite{emmanouilidou2008electron,Arkhipov2019_OL, arkhipov2020some, arkhipov2020selective, arkhipov2021population,Pakhomov_2022}, acceleration of charged particles ~\cite{rosanov2020direct}, coherent state transfer and ionization of Rydberg atoms \cite{Bucksbaum, Mestayer} or steering the spin dynamics \cite{aleksandrov2020relativistic, rosanov2021electron}.

Compared to conventional harmonic few-cycle pulses, unipolar pulses have no carrier frequency and possess broad spectrum from zero frequency up to visible range, which allows faster data transmission and processing \cite{arkhipov2023unipolar, diachkova2023light}. Besides the unipolar subcycle pulses can be used for the holographic recording with ultrahigh temporal resolution \cite{arkhipov2020possibility}, creation of population density gratings \cite{arkhipov2021population} and formation and control of dynamical cavities in resonant media \cite{diachkova2023light}.

A difference between the action of subcycle and single cycle optical attosecond pulse on atomic systems was observed experimentally~\cite{Hassan}.
Furthermore, in the case of extremely short pulses, $\tau_p < T_g$, the standard theoretical formalism such as Keldysh theory becomes improper for the description of the light-matter interaction~\cite{Dimitrovski_PRL,Dimitrovski_PRA,rosanov2021criterion}. 
For such a case, new physical quantities should be introduced and standard approaches as well as phenomena in optics studied for long multi-cycle pulses should be revised as follows. 
The impact of the half-cycle pulses on micro-object is determined by the electric pulse area $S_{E}$ rather than the pulse energy or its amplitude~\cite{arkhipov2021atomic, rosanov2021criterion}. Besides when speaking about subcycle pulses their direct interference is impossible. And their impact on microobjects can be described in terms of their area interference \cite{arkhipov2021envelope,arkhipov2022interference}. We remark that for the control of quantum systems it is not necessary to obtain strictly unipolar pulse, but pulses can contain a weak trailing tail of the opposite polarity, thus the total electric area can be zero. In this case, the pulse action on atomic system is primarily determined by the electric area of the main half-wave \cite{arkhipov2020selective}.

To assess the efficiency of the extremely short pulse impact on different quantum systems, the electric pulse area should be compared with its characteristic scale, which can be introduced for arbitrary quantum system~\cite{arkhipov2021atomic}. The ultrashort pulse transfers the kinetic momentum $\delta\vec{p}=q\vec{S_E}$, where $q$ is the electric charge, to the medium. For the quantum system with the characteristic size of the electron localization, $a$, the eigenmomentum can be estimated from the uncertainty principle $p_0\sim\hbar/a$.
That is a new quantity, the atomic scale of the electric pulse area of the quantum system, $S_{{QW}}\sim\frac{\hbar}{qa}$ was recently introduced~\cite{arkhipov2021atomic, rosanov2021criterion}. It is inversely proportional to the characteristic size $a$ of the system and allows one to assess the efficiency of the unipolar and quasi-unipolar pulse effect on different quantum systems.

Alongside the attosecond physics, an active investigation of the optical properties of nanostructures (metallic and dielectric nanoparticles, semiconductor quantum dots, etc.) led to a significant progress in nanophysics. Such objects of nanometer sizes have discrete atomic-like structure of energy levels, which makes them to be remarkable tools in various applications \cite{scholl2012quantum,kuznetsov2016optically, smirnova2016multipolar,carletti2017controlling,lepeshov2019hybrid,seiffert2022strong}. In particular, metallic nanoparticles or nanoantennas enable to tune the light properties at the nanoscale \cite{chen2012enhanced, biagioni2012nanoantennas, lepeshov2019hybrid, smirnova2016multipolar}. Nanoantennas can be used for creation of single-photon sources \cite{curto2010unidirectional, busson2012accelerated} and nanolasers \cite{zhang2015plasmonic, gongora2017anapole}. Plasmonic resonances in metallic nanoparticles are of considerable interest in nanooptics for realization of compact quantum devices, including single-photon sources and transistors~\cite{scholl2012quantum,saha2012gold,tame2013quantum}. Besides, the photoelectron emission from nanostructures can be used for generation of ultrafast switchable on-chip electronic current \cite{shi2021femtosecond,lovasz2022nonadiabatic}.

In general, the semiconductor heterostructures allow for a significant flexibility of the radiative properties \cite{Alferov1998,Ivchenko,BELOV201996}. Dielectric nanoparticles and metasurfaces can be used to enhance the second-harmonic generation as well as the electric and magnetic near-fields~\cite{kuznetsov2016optically,carletti2017controlling}. Similarly, the semiconductor quantum dots and quantum wells (QWs) are the active media and saturable absorbers in the mode-locked semiconductor lasers~\cite{Rafailov,yadav2023edge}. Recently these two well-developed areas of modern physics - the attosecond physics and nanophysics have become close to each other and have been combined into the novel area - ``attonanophysics'', see review~\cite{ciappina2017attosecond,ciappina2023multiphoton} and references therein. The interaction of few-cycle pulses with nanostructures is becoming a relevant topic of modern optics now \cite{seiffert2022strong}. However, the interaction of half-cycle unipolar and quasi-unipolar pulses with nanostructures is poorly studied so far.

In this paper, we theoretically study the interaction of a strong half-cycle pulse with a QW nanostructure via approximate analytical and direct numerical solution of the time-dependent Schr{\"o}dinger equation. The model of the one-dimensional rectangular QW with infinite and finite barriers is considered. It is shown that, when the incident pulse duration is smaller than $T_g$, the population dynamics of the electron in such a QW depends on the ratio of the electric pulse area to its characteristic measure $S_E/S_{{QW}}$. The latter is inversely proportional to the QW width. A possibility of the selective control of quantum level population and the creation of the population inversion by subcycle pulses is demonstrated. These results reveal the possibility of the effective and selective ultrafast control of the population dynamics of electrons in QWs.

\section{Interaction of a subcycle attosecond pulse with QW nanostructures: theoretical approach}

The interaction of an incident pulse with a  quantum system can be described by the time-dependent Schr{\"o}dinger equation (TDSE) for the wave function $\Psi (\vec r, t)$ ~\cite{Landau}:
\begin{equation}
i \hbar \ \frac{\partial \Psi}{\partial t} = \Big[ \hat H_0 + \hat V(t) \Big] \Psi.
\label{SE}
\end{equation}
Here $\hat H_0$ is the intrinsic Hamiltonian of the quantum system, which includes the time-independent potential $V(\vec r)$. The time-dependent part is contained in the operator term $\hat V(t) = - q \vec r \vec E(t)$  representing the dipole approximation of the interaction of an electron of charge $q$ with an external electric field $\vec E(t)$.

Nowadays, the subcycle attosecond pulses have durations comparable or even shorter than the periods of resonant transitions in quantum systems. We thus assume that the driving pulse duration $\tau_p$ is much shorter than the characteristic time of the internal dynamics of the considered quantum system, e.g. the oscillation period of the ground state $T_g$:
\begin{equation}
\tau_p  \ll  T_g.
\label{sudden}
\end{equation}
In this case, TDSE can be solved approximately using the so-called theory of sudden perturbations introduced by Migdal ~\cite{migdal1939ionizatsiya,Landau} (see also  \cite{dykhne1978jarring, bohm2012quantum, Dimitrovski_PRL, Dimitrovski_PRA}).

Let us assume that before the pulse the quantum system is in the ground state with the wave function $\Psi (\vec r, 0) = \Psi_{1} (\vec r)$. Then, in the first approximation, one can neglect in TDSE Eq.~\eqref{SE} the term $\hat H_{0}$ during the action of the pulse, as compared to the term $\hat V(t)$, since according to Eq.~\eqref{sudden} ~\cite{Dimitrovski_PRA, Rosanov2018, Arkhipov2019_OL}:
\begin{eqnarray}
\nonumber
\Psi (\vec r, \tau_p) &=& \Psi (\vec r, 0) \exp{\Big( -\frac{i}{\hbar} \int_{0}^{\tau_p} [{\hat{H}}_{0} + \hat V(t)] dt \Big)}   \\
\nonumber
&=& \Psi (\vec r, 0) \exp{\Big( -\frac{i}{\hbar} \int_{0}^{\tau_p} \hat{H}_{0} dt \Big)} \times \\
\nonumber
&&  \exp{\Big( -\frac{i}{\hbar} \int_{0}^{\tau_p} \hat V(t) dt \Big)}  \\
&\approx&  \Psi_1 (\vec r) \ \exp{\Big( \frac{i}{\hbar} q \vec r \vec S_E \Big)}, \ \ \ \ 
\label{sudden_solution}
\end{eqnarray}
where the first exponential factor has the index of power of the order of $\tau_p / T_g$ and is thus very close to unity due to the condition Eq.~\eqref{sudden}. The resulting wave function after the end of the pulse $\Psi (\vec r, \tau_p)$ therefore solely depends on the electric area of the incident pulse $\vec S_{E}$ ~\cite{Rosanov2018}. Physically neglecting the term $\hat H_{0}$ assumes that pulse is so short that the electron position (e.g. in classical Bohr orbit) does not change significantly during the pulse action. This is equivalent to neglecting the atomic potential during the pulse action. Such approximation is used in the problems of atom "jolt" by a short-term perturbation (see problems 2 and 3 after par. 43, Ch. 6 in \cite{Landau} and review  \cite{dykhne1978jarring}) and in the study of atomic ionization by strong half-cycle pulses \cite{Dimitrovski_PRL, Dimitrovski_PRA,briggs2008ionization,grozdanov2009model,lugovskoy2015sudden,Rosanov2018,Arkhipov2019_OL,rosanov2021criterion}.

The wave function of the system $\Psi (\vec r, \tau_p)$ can be then expanded in terms of the eigenfunctions $\Psi_n (\vec r)$:
\begin{equation}
\nonumber
\Psi (\vec r, \tau_p) = \sum_n a_{n} \Psi_n(\vec r),
\end{equation}
to yield the amplitudes of the eigenstates as:
\begin{equation}
a_n = \int \Psi(\vec r, \tau_p) \Psi_n(\vec r) d\vec r.
\label{an}   
\end{equation}

We would like to particularly emphasize that the sudden-perturbation approximation provided by Eqs.~\eqref{sudden}-\eqref{sudden_solution} should not be confused with the standard perturbation theory. The standard perturbation theory is valid, when the external driving field is much weaker than the internal field in the system. The ratio of the strengths of the external and the internal fields then acts as a small parameter for the approximate solution of the time-dependent Schr{\"o}dinger equation Eq.~\eqref{SE}.

In contrast, the sudden-perturbation approximation is valid as long as the duration of the driving pulse is much smaller than the characteristic time of the internal dynamics of the quantum system. In this case it is the ratio of the exciting pulse duration and the characteristic time of the internal dynamics of the quantum system that acts as a small parameter for the approximate solution of the Schr{\"o}dinger equation Eq.~\eqref{SE}. At the same time, the strength of the external field is in most cases not relevant for the validity of the sudden-perturbation approximation, so that it could be both stronger and weaker than the internal field in the system. In the following sections this approach Eqs.~\eqref{sudden}-\eqref{sudden_solution} is used for the calculation of the bound state populations in QWs.

\subsection{Narrow QW with finite barriers}

\begin{figure}[htpb]
\centering
\includegraphics[width=1.1\linewidth]{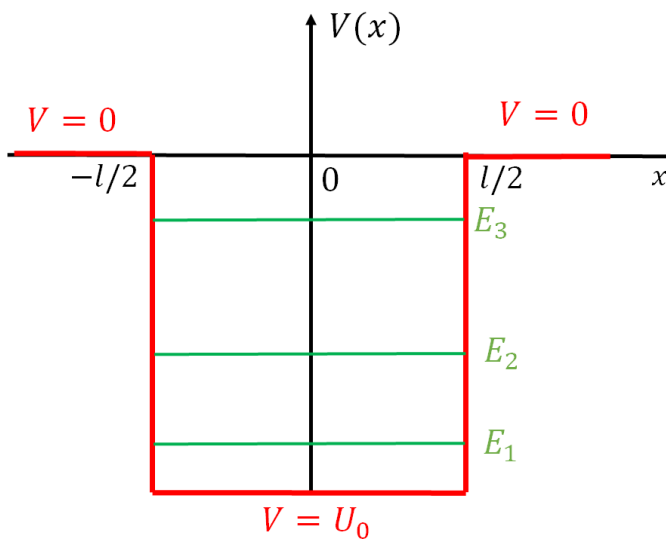}
\caption{(Color online) The scheme of the QW with finite barriers. }
\label{figFiniteQW}
\end{figure}

Let us consider theoretically a relatively narrow one-dimensional rectangular QW having only one bound state. The time-independent potential $V(x)$ is defined as, see Fig.~\ref{figFiniteQW}:
\begin{equation}
\label{U0}
V(x) = \left\{
  \begin{array}{lr}
    0 & \mbox{ if  } |x| > l/2 \\
    U_{0} & \mbox{ if  } |x| < l/2
  \end{array}
\right. .
\end{equation}
For simplicity of the analytical solution we consider a narrow QW possessing only one bound state. Thus, we look for the sole even solution. For $|x| \le l/2$, $\Psi=\alpha_{1}\cos kx$, where $k^2=2m(E - U_0)/\hbar^2$ and $\alpha_{1}$ is the normalization factor. For $|x| > l/2$,  $\Psi=\alpha_2 e^{-\kappa x}$, where $\kappa^2 = -2mE/\hbar^2$. From the continuity condition for $\Psi'/\Psi$ at the $x=l/2$, one obtains the transcendental equation for the energy levels~\cite{Landau}
$\sqrt{y} \tan(\mu\sqrt{y})=\sqrt{1-y}$, where $y = 1 - E/U_0$ and $\mu=l\sqrt{-m\,U_0/2}/\hbar$.

From the transcendental equation, one can see that the case of narrow QW with a single bound state of the energy $E \approx 0$ implies $\mu<<1$. As shown in Ref.~\cite{Landau}, in this case, the energy of the bound state can be approximated by expansion  $E = -\frac{m l^{2}}{2\hbar^2}U_0^2$. 
From the continuity of wave functions, one can find that $\alpha_2=\cos(kl/2)\exp(\kappa l/2)\alpha_1$.
From the normalization condition $\int |\Psi(x)|^2 dx=1$ it is easy to obtain 
\begin{equation}
\alpha_1 = \left(l/2 + \sin(kl)/2k + \cos^{2} (kl/2)/\kappa \right)^{-1/2}.
\end{equation}
Thus, from Eq.~(\ref{an}) we find for the amplitude of the single bound state \cite{arkhipovQW2023e}:

\begin{eqnarray}
\nonumber
a_1= \frac{\sin\left( \frac{S_E}{S_{{QW}}}  \right) }{ \frac{qS_E}{\hbar}} + \frac{\sin\left( kl+ \frac{S_E}{S_{{QW}_{at}}}\right)}{ 4k + \frac{2qS_E}{\hbar}} + \\
\nonumber
\frac{\sin\left( kl- \frac{S_E}{S_{{QW}}} \right) }{ 4k-\frac{2qS_E}{\hbar}} + 2 Re \frac{\left(-\kappa l + il\frac{S_E}{S_{{QW}}}   \right)}{\kappa-i\frac{S_E}{S_{{QW}}}}.
\label{w0}
\end{eqnarray}

The population of the bound state after the pulse is given by $w_1=|a_1|^2$ and corresponding ionization probability  $w_{ion}=1-w_1$. From Eq.~(\ref{w0}), it is seen that it is proportional to the ratio of the electric pulse area $S_E$ to its corresponding atomic scale of the QW system $S_E/S_{{QW}}$, where  $S_{{QW}}=\frac{2\hbar}{q\, l}$ \cite{arkhipov2021atomic,arkhipov2022control}. The population of the single bound state decays very fast with an increase of the electric pulse area. When $S_{E}=0$, the QW remains unexcited: $w_{0}=1$. This case strongly differs from the well-known multi-cycle pulse excitation~\cite{seiffert2022strong}.

We remark that the model of the QW with a single bound state is used for the description of the deuteron ~\cite{bohm2012quantum} This QW has the radius $r_d=2.8\cdot10^{-13}$~cm and the binding energy is $E_1=2.23$~MeV. The corresponding time is $T_g=0.0018$~as. For deuteron, the corresponding atomic scale of electric pulse area is given as $S_{D}=2\hbar/qr_d \approx 10^{-10}$ erg$\cdot$sec/ESU \cite{arkhipovQW2023e}. Thus, for the effective excitation and control of the deuteron the half-cycle gamma and X-ray attosecond pulses with large electric area are needed.

It is interesting to consider as a limiting case a very deep ($U_0 \to - \infty$)  and very narrow ($l \to 0$) rectangular potential well such that the area of this well $U_0 l$ remains constant. In this case, the so-called zero-radius potential $V(x)=-V_0\delta(x)$ is obtained, where $\delta(x)$ is the Dirac delta-function. Quite similar results can be obtained for such the potential. In particular, this well contains also a  single bound state $E=-\frac{m}{2\hbar^2}V_0^2$ \cite{zettili2003quantum}. Such potential is used in different problems of atomic physics and for modeling the interaction of ions with ultra-short light pulses \cite{demkov2013zero,grozdanov2009model}.  In this case, using the approach described above it is easy  to show that the probability to remain in the bound state is given by (see \cite{arkhipovion}):
\begin{equation}
\nonumber
w_0=\frac{1}{\left(1+\frac{S_E^2}{S_0^2}\right)^2},
\end{equation}
where $S_0=\hbar/qx_0$ is the corresponding ion scale of the electric pulse area, $x_0$ is the electron localization size in the bound state.

\subsection{Rectangular QW with infinite barriers}
\begin{figure}[htpb]
\centering
\includegraphics[width=1.1\linewidth]{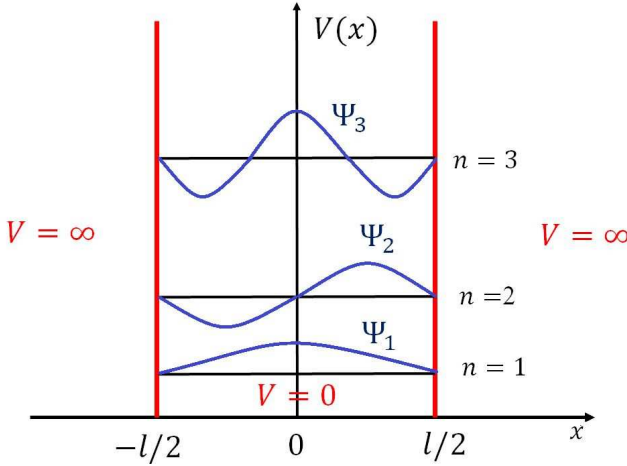}
\caption{(Color online) The scheme of the QW with infinite barriers. A few lowest eigenfunctions are depicted.}
\label{fig0}
\end{figure}

Let us consider the rectangular QW with the infinite barriers, i.e. the potential
\begin{equation}
\label{U0}
V(x) = \left\{
  \begin{array}{lr}
    0 & \mbox{ if  } |x| < l/2 \\
    \infty & \mbox{ if  } |x| \ge l/2
  \end{array}
\right. ,
\end{equation}
see Fig.~\ref{fig0}.

The corresponding wave functions are given by~\cite{Landau}:
\begin{eqnarray}
\Psi_n(x) = \sqrt{\frac{2}{l}} \cos\left(\frac{\pi n}{l} x \right), n= 1,3,5 \ldots \\
\Psi_n(x) = \sqrt{\frac{2}{l}} \sin\left(\frac{\pi n}{l} x \right), n= 2,4,6 \ldots
\label{wf}
\end{eqnarray}
and the energy levels are \cite{Landau}:
\begin{equation}
\label{EnergyQW}
 E_n=\frac{\hbar^2}{2m}\left(\frac{\pi n}{l}\right)^2.   
\end{equation}
Using these expressions and equations Eqs.~\eqref{sudden_solution},~\eqref{an} one can obtain for the amplitudes of the odd states: 
\begin{eqnarray}
\nonumber
a_n=\frac{2}{l}\int \Psi_n(x) \Psi_1(x) \exp{\Big( \frac{i}{\hbar} qxS_E \Big)} dx=\\
\nonumber
\frac{\sin\left( \frac{\pi(n+1)}{2} +\frac{S_E}{S_{{QW}}} \right) }{\pi(n+1) + 2\frac{S_E}{S_{{QW}}}} + \frac{\sin\left( \frac{-\pi(n+1)}{2}  +\frac{S_E}{S_{{QW}} } \right) }{-\pi(n+1) + 2\frac{S_E}{S_{{QW}} }} +  \\
\nonumber
\frac{\sin\left( \frac{\pi(n-1)}{2} +\frac{S_E}{S_{{QW}}} \right) }{\pi(n-1) + 2\frac{S_E}{S_{{QW}}} } + \frac{\sin\left( \frac{\pi(1-n)}{2} +\frac{S_E}{S_{{QW}}} \right) }{\pi(1-n) + 2\frac{S_E}{S_{{QW}} }}.  \\
\label{amp_odd}
\end{eqnarray}

In the same way, for the amplitudes of the even states we obtain:
\begin{eqnarray}
\nonumber
 a_n=\frac{2}{l}\int \Psi_n(x) \Psi_1(x) \exp{\Big( \frac{i}{\hbar} qxS_E \Big)}dx=\\
\nonumber
- i\frac{\sin\left( \frac{\pi(n+1)}{2} +\frac{S_E}{S_{{QW}_{at}}   } \right) }{\pi(n+1) + 2\frac{S_E}{S_{{QW}_{at}} }} + i\frac{\sin\left( \frac{-\pi(n+1)}{2} +\frac{S_E}{S_{{QW}_{at}} } \right) }{-\pi(n+1) + 2\frac{S_E }{S_{{QW}_{at}} }} - \\
\nonumber
i\frac{\sin\left( \frac{\pi(n-1)}{2} +\frac{S_E }{S_{{QW}_{at}} } \right) }{\pi(n-1) + 2\frac{S_E }{S_{{QW}_{at}}}} + 
 i\frac{\sin\left( \frac{\pi(1-n)}{2}  + \frac{S_E}{S_{{QW}_{at}}} \right) }{\pi(1-n) + 2\frac{S_E}{S_{{QW}_{at}}}}. \\ 
 \label{amp_even}
 \end{eqnarray} 
One can see that populations of odd and even states depend on the pulse area $S_{E}$ divided by its atomic scale $S_{{QW}}=2\hbar/ql$. If the driving pulse area $S_E$ decreases to $0$, the population of the ground state tends to $1$, while for upper states $n>1$ the populations vanish: $a_{n} \to 0$. That means that under the impact of a zero-area single-cycle pulse the system remains in the ground state. This fact can be confirmed by the direct calculations using Eqs.~(\ref{amp_odd})-(\ref{amp_even}).

\section{Results of numerical calculations and discussion}

We proceed with performing the numerical simulation of the spatio-temporal behaviour of a one-dimensional rectangular QW excited by a short unipolar pulse. To this end, the finite-difference discretization of the time-dependent Schr\"{o}dinger equation was employed. Over the time variable, the absolutely stable Crank-Nicolson scheme was realized, which assures the second-order uncertainty of the numerical solution both in space and time $\sim (\Delta x)^2 + (\Delta t)^2$. The rectangular numerical grid with the spatial grid spacing $\Delta x = 0.005$~nm and the temporal spacing $\Delta t = 2$~attosec was used.

For quantum wells with infinite barriers we applied zero boundary conditions for the wave function. For quantum wells with finite barriers the absorbing boundary conditions were implemented using the complex-scaling technique ~\cite{MOISEYEV1998212,PhysRevB.105.155417}. 
Such approach is based on the scaling of the coordinate variables into the complex plane and provides the damping of the outgoing waves outside the QW by generating the exponentially vanishing functions, while keeping the interaction region unperturbed and, thus, does not change the dynamics of the quantum system. As the result, one can simulate the wave function asymptotic propagation with transparent boundaries in a finite computational domain. The more detailed description of this numerical method and its validity test can be found in ~\cite{belov2023formation}.

\subsection{Rectangular QW with infinite barriers}

\begin{figure}[htpb]
\centering
\includegraphics[width=1.0\linewidth]{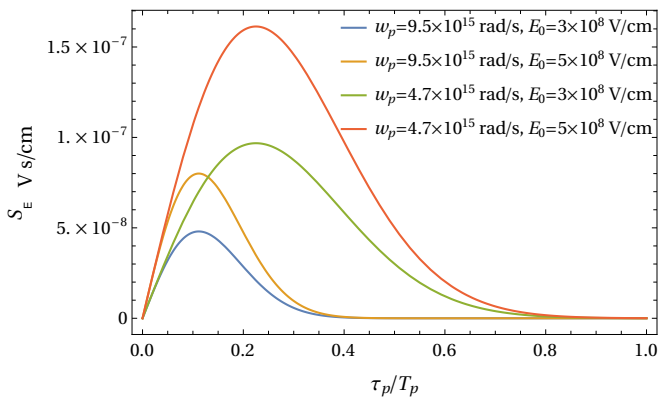}
\caption{(Color online) Dependence of the electric pulse area $S_{E}$ on the pulse duration $\tau_p$ in units of $T_{p}$ for different parameters of the pulse, CEP is taken to be $\vartheta = 0$.}
\label{fig01}
\end{figure}

Here, we consider a one-dimensional rectangular QW with infinite barriers, see Fig.~\ref{fig0}. The width of the QW is set to be $l=1.2$~nm.
The energy of the ground state is defined as
$E_{1}=\hbar^2 \pi^2/2ml^2=259$~meV, that is $4.14\cdot10^{-13}$~erg.
The energies of upper states can also be easily obtained by Eq.~(\ref{EnergyQW}).
Then, for example, the corresponding transition wavelengths are $\lambda_{12}=1600$~nm, $\lambda_{13}=600.33$~nm, $\lambda_{14}=320.17$~nm, which are typical for metallic nanostructures. The driving attosecond pulse is considered in the form
$$
E(t) = E_0 \exp{(-t^2/\tau_p^2)} \cos(\omega_p t + \vartheta),
$$
where $E_0$ is the pulse amplitude, $\tau_p$ is the pulse duration. Since ultraviolet and optical attosecond pulses have become available now~\cite{Krausz, Wu, Xu, Hassan}, in our simulations we assume $\omega_p$ of the order of $4.7\cdot10^{15}$ rad/s ($\lambda_p = 400$~nm, $T_p=2\pi/\omega_p=1.33$~fs), which is in the ultraviolet frequency range. The electric pulse area $S_{E}$ is given as $S_{E} = E_{0} \tau_{p} \sqrt{\pi} e^{-\omega_{p}^2 \tau_{p}^2 /4} \cos \vartheta$. It depends on the pulse duration and the carrier-envelope phase (CEP). In the following, the CEP is taken to be $\vartheta = 0$. Fig.~\ref{fig01} shows the pulse area $S_E$ vs. the pulse duration $\tau_p$ for different values of the pulse amplitude $E_0$ and the parameter $\omega_p$. The obtained curve reaches its maximum  for $\tau_p$ around $0.2 T_{p}$, while in the limit of a single-cycle pulse with $\tau_p/T_p = 1$ the  electric pulse area goes to zero $S_E \to 0$.

\begin{figure}[htpb]
\centering
\includegraphics[width=1.0\linewidth]{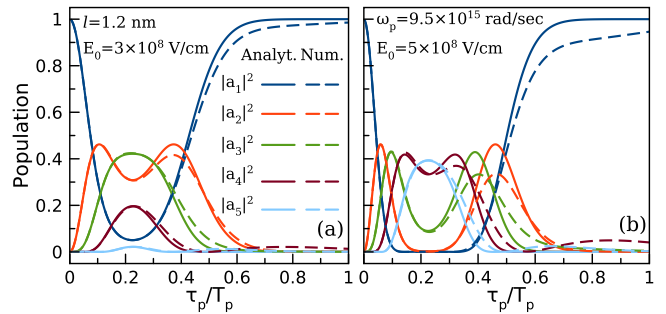}
\caption{(Color online) Dependence of the populations ($|a_1|^2, ..., |a_5|^2$) on the pulse duration $\tau_p/T_{p}$. The analytically as well as the numerically obtained results are plotted. The values of the parameters are denoted. The parameter CEP $\vartheta = 0$.}
\label{fig1}
\end{figure}

Let us consider a numerical example. Fig.~\ref{fig1} shows the populations of five lowest quantum states ($|a_1|^2, ..., |a_5|^2$) in the considered rectangular QW as a function of the driving pulse duration $\tau_p$ for $\omega_{p}=9.5\cdot10^{15}$ rad/sec for two pulse amplitudes: $a)$ $E_0=3\cdot10^8$ V/cm and $b)$ $E_0=5\cdot10^8$ V/cm. The solid lines show values calculated by means of the analytical expressions (\ref{amp_odd}-\ref{amp_even}). Let us consider the left plot Fig.~\ref{fig1} $(a)$. The dependencies can be easily understood based on the correspondent analytical electric pulse area curves in Fig.~\ref{fig01}. Since the pulse area increases monotonically with the increase of $\tau_p$ until its maximum, the population of the ground state decreases, whereas populations of the excited states grow. When $S_E$ approaches the maximum, the population of the ground state becomes almost zero, but the upper levels appear significantly more populated. With further increase of $\tau_p$ the electric pulse area decreases and populations of the upper levels decrease. The ground state population eventually recovers to the initial unity value.

\begin{figure}[htpb]
\centering
\includegraphics[width=1.0\linewidth]{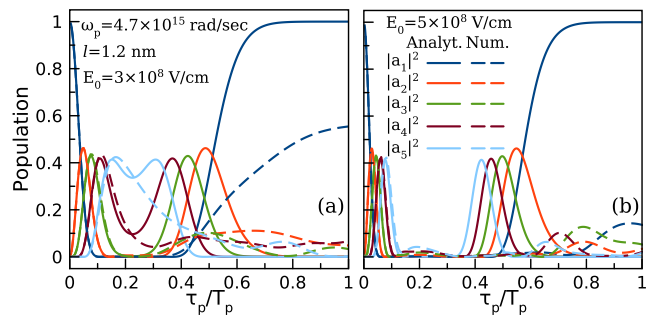}
\caption{(Color online) Dependence of the populations ($|a_1|^2, ..., |a_5|^2$) on the pulse duration $\tau_p/T_{p}$. The analytically as well as the numerically obtained results are plotted. The disagreement between analytical and numerical results originates from the fact that $w_{p}$ is close to the QW eigenfrequency. The values of the parameters are denoted. The parameter CEP $\vartheta = 0$.}
\label{fig2}
\end{figure}

\begin{figure}[htbp]
\centering
\includegraphics[width=1.0\linewidth]{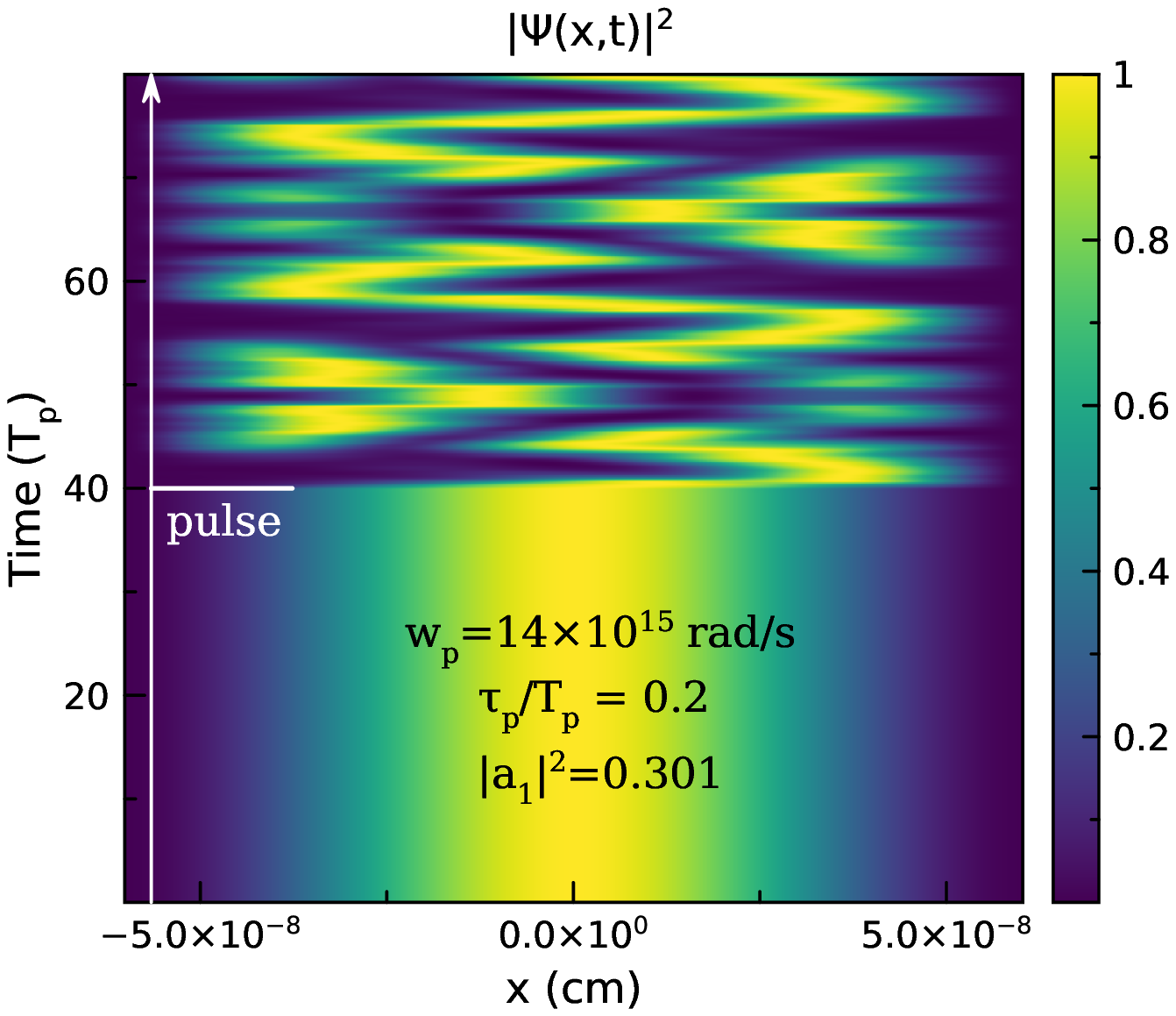}
\caption{The wave packet propagation in the one-dimensional QW with infinite barriers. The ratio $\tau_{p}/T_{p}=0.2$. A magnitude of the wave function is normalized to unity for convenience.}
\label{fiWP02}
\end{figure}
\begin{figure}[htbp]
\centering
\includegraphics[width=1.0\linewidth]{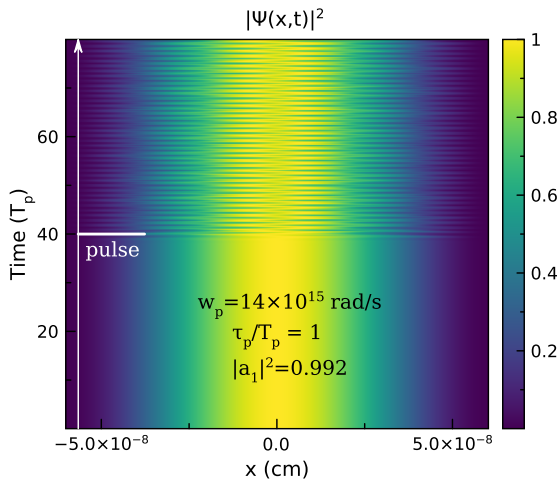}
\caption{The wave packet propagation in the one-dimensional QW with infinite barriers. The ratio $\tau_{p}/T_{p}=1$. A magnitude of the wave function is normalized to unity for convenience.}
\label{figWP1}
\end{figure}

This maximum shows the possibility of the selective control of the quantum level populations in the QW. For the selective control of populations in atoms resonant multi-cycle pulses can be used. In metals it is impossible because of low relaxation times. The results of  Fig.~\ref{fig1} clearly demonstrate the possibility of the selective control of the population of quantum levels. We remark that the similar selectivity produced by a unipolar pulse train was demonstrated earlier in atomic system like the hydrogen atom, etc.~\cite{Arkhipov2020_OS, arkhipov2020selective}. When the pulse duration approaches $T_p$, the pulse area is almost zero and the population of the ground state tends to $1$, while other levels become unpopulated.

The dashed curves in Fig.~\ref{fig1} show our results based on the numerical solution of the TDSE~(\ref{SE}). The calculations were performed using the Crank-Nicolson method~\cite{crank_nicolson_1947}. A more detailed description of the numerical method used in the calculations as well as numerical algorithm test results  can be found in the Appendix presented in Ref. \cite{belov2023formation}.
One can see that some correspondent solid and dashed curves visually overlap, thus a good agreement between the analytical and numerical results takes place.
In Fig.~\ref{fig2} we show the same dependencies but for $\omega_p=4.712\cdot10^{15}$ rad/s. One can see that in these plots the numerical results significantly differ from the analytical ones. This disagreement originates from the fact that the parameter $\omega_{p}$ is of the same order as the eigenfrequency of the ground state. As it was already described, the developed theory is appropriate only when $T_{p} \ll T_{g}$.

Figs.~\ref{fig1} and~\ref{fig2} clearly show an example of the population inversion, which can be realized in the QW structure. Indeed, when ($\tau_p\sim0.012T_p$)
the population of the fourth state $|a_2|^2\sim0.4$ (magenta curve in Fig.~\ref{fig2} $(a)$), whereas the population of the ground level and other levels are significantly smaller. In QW the resonant transitions $1-2$ and $2-3$ are allowed. At the same time, the transition $1-3$ is forbidden. These circumstances suggest using the subcycle pulses for creation of the population inversion in QWs and lasing. We remark that QW nanostructures were suggested for lasing in the so-called plasmon nanolasers or spasers \cite{bergman2003surface, gwo2016semiconductor,
premaratne2017theory, balykin2018plasmon}. The spaser is a very small laser having the subwavelength dimensions and a low-Q plasmonic resonator, whose operation principle is based on the plasmonic oscillations using a stimulated emission of the surface plasmons \cite{premaratne2017theory}.

Figs.~\ref{fiWP02}-\ref{figWP1} show the corresponding wave packet evolution for the excitation by a half-cycle pulse with the ratio $\tau_{p}/T_{p}=0.2$ and by a single-cycle pulse with $\tau_{p}/T_{p}=1$, respectively. The excitation pulse arrives at the time point $t = 40 T_p$. We would like to recall here that we choose  the driving pulse duration $\tau_p$ to be much shorter than the characteristic time of the internal dynamics of the considered quantum system, e.g. the oscillation period of the ground state $T_g$. This means that the exciting pulse is so short that it is not able to cause any observable displacement of the wave function inside the QW over the entire pulse duration. Still the driving pulse can transfer a certain momentum to the bounded electron. That is why no internal dynamics of the wave function can be seen during the pulse action in figures like Figs.~\ref{fiWP02}-\ref{figWP1}. Instead, the pulse-induced dynamics is only to be visible over a time slot of the order of $T_g$, i.e. on a much larger time scale as compared to the pulse duration $\tau_p$. Therefore we plot the wave function evolution in Figs.~\ref{fiWP02}-\ref{figWP1} across the time interval of several tens of $T_g$ in duration to provide the proper illustration of the wave function evolution in time.

In Fig.~\ref{fiWP02} it is seen that the half-cycle pulse destroys the ground state, while creating a superpositon of basic eigenstates after the pulse, see complex oscillations in the top of Fig.~\ref{fiWP02}. At the same time, for the single-cycle pulse with $\tau_{p}=T_{p}$ in Fig.~\ref{figWP1}, the ground state population is close to unity after the pulse ending. An observed small perturbation originates from multiple reflections from the infinite QW barriers. As a result, in average the corresponding wave packet shape resembles the ground state one, see Fig.~\ref{figWP1}.

\begin{figure}[htbp]
\centering
\includegraphics[width=1.0\linewidth]{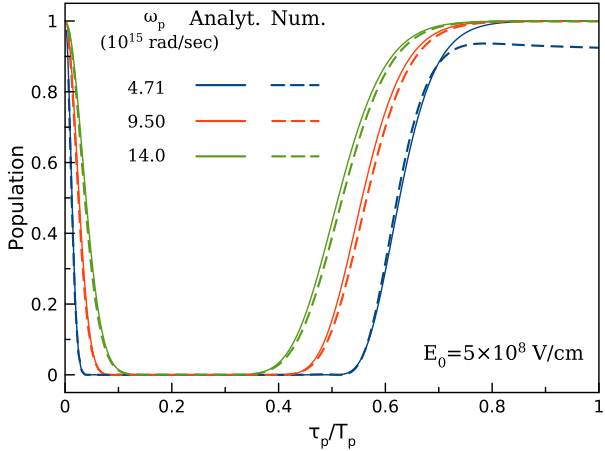}
\caption{The population of the single bound state in the finite QW during the excitation by an incident pulse with the amplitude $E_{0}=5\times 10^{8}$~V/cm. The solid lines are the analytical results, whereas the dashed ones are the numerical data. The potential $V_{0}=0.2$~eV.}
\label{figFin5V}
\end{figure}

\begin{figure}[htbp]
\centering
\includegraphics[width=1.0\linewidth]{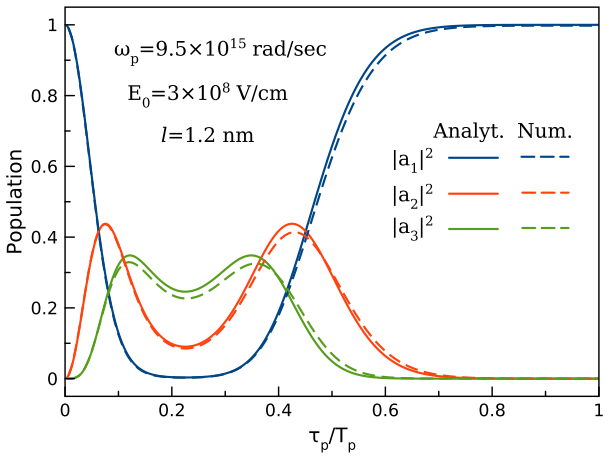}
\caption{Analytically and numerically obtained populations of the three bound states in the finite QW during the excitation by an incident pulse with the amplitude $E_{0}=3\times 10^{8}$~V/cm. The analytical results are based on the theory of sudden perturbations, taking into account the numerical wavefunctions of a particle in the finite QW. The potential $V_{0}=1.5$~eV.}
\label{figFin3levels}
\end{figure}

\subsection{Rectangular QW with finite barriers}


Let us consider the numerical simulation of the excitation of an open quantum system having a single bound state. For the rectangular QW of width $l=1.2$~nm, it can be obtained, for instance, by taking the barrier height to be $V_{0}=0.2$~eV. The Fig.~\ref{figFin5V} shows the population of the single bound state in a rectangular QW excited by the unipolar short pulse as a function of the ratio $\tau_{p}/T_{p}$ characterising the relative pulse duration. The dashed curves present results of numerical modeling, whereas the solid ones are analytically obtained data based on formulas of the previous section. One can see that the numerical and analytical results are in good agreement. The very different populations for different values of $\tau_{p}$ imply the possibility of the effective control of the quantum system. A notable difference of the population of the state excited by the pulse of $\omega_{p}=4.712\cdot 10^{15}$rad/s takes place because this frequency is more close to a frequency of the transition to the continuum.

Fig.~\ref{figFin3levels} shows the analytical and numerical populations for the finite QW with three bound states. Such a configuration is achieved by the barrier height of 1.5~eV. The system is excited by a short unipolar pulse with $\omega_{p}=9.5\times 10^{15}$~rad/s. One can see a good agreement between numerical and analytical data for this case.

\begin{figure}[htbp]
\centering
\includegraphics[width=1.0\linewidth]{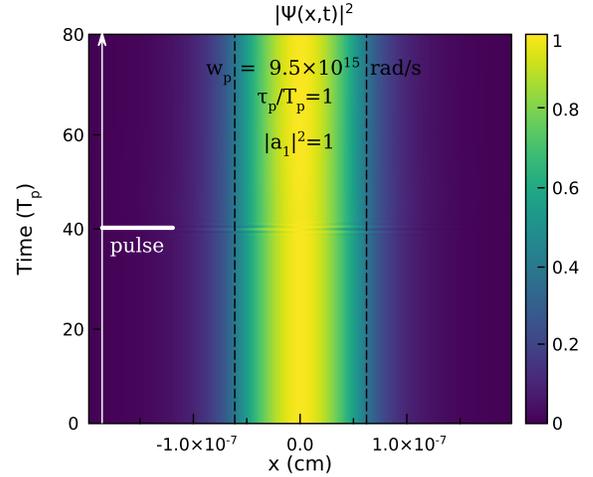}
\caption{The wave packet propagation in the one-dimensional QW with finite barriers of 0.2~eV. The ratio $\tau_{p}/T_{p}=1$. The dashed lines show the QW barriers. A magnitude of the wave function is normalized to unity for convenience.}
\label{figFinWP1}
\end{figure}

\begin{figure}[htbp]
\centering
\includegraphics[width=1.0\linewidth]{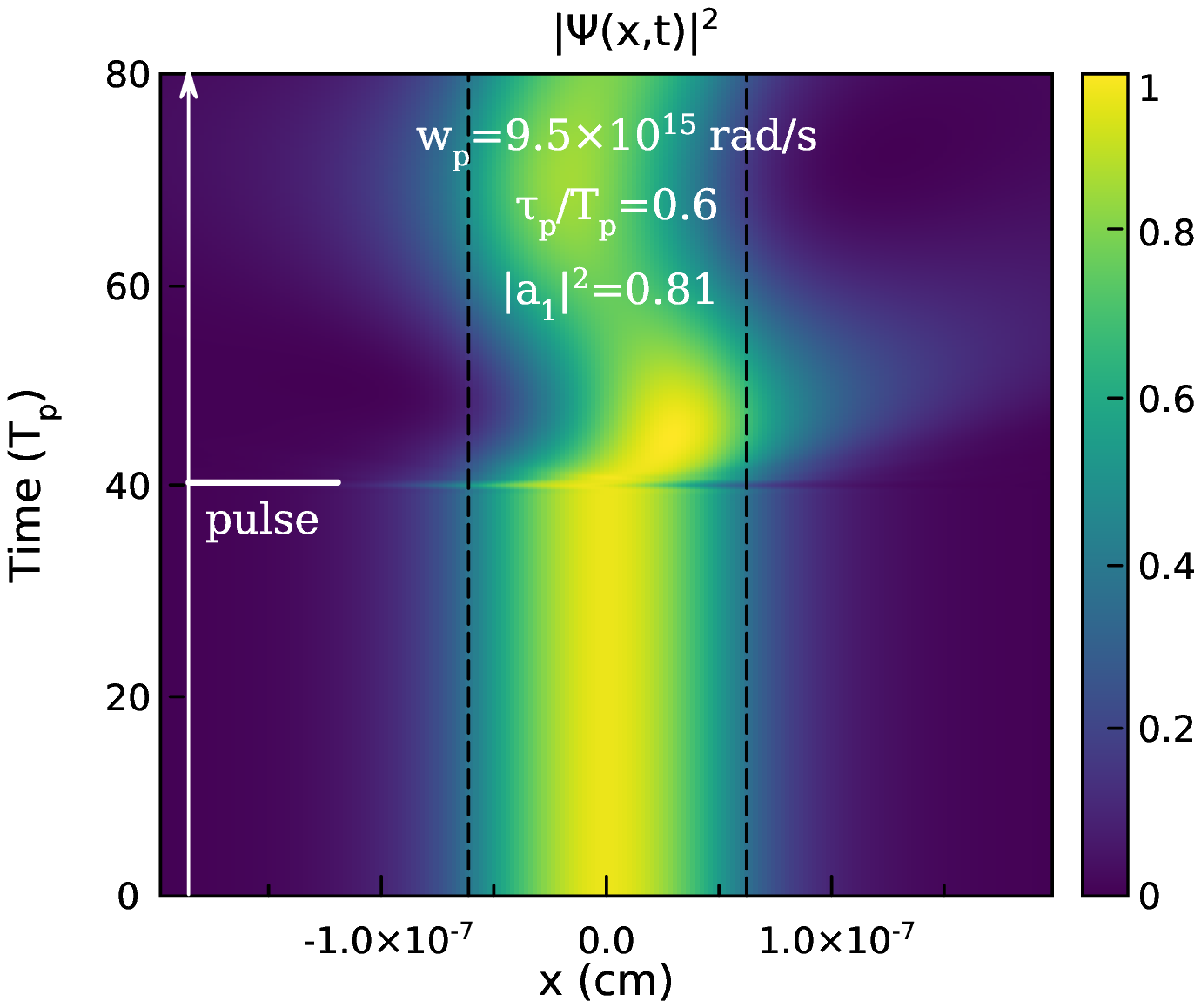}
\caption{The wave packet propagation in the one-dimensional QW with finite barriers of 0.2~eV. The ratio $\tau_{p}/T_{p}=0.6$. The dashed lines show the QW barriers. A magnitude of the wave function is normalized to unity for convenience.}
\label{figFinWP06}
\end{figure}

\begin{figure}[htbp]
\centering
\includegraphics[width=1.0\linewidth]{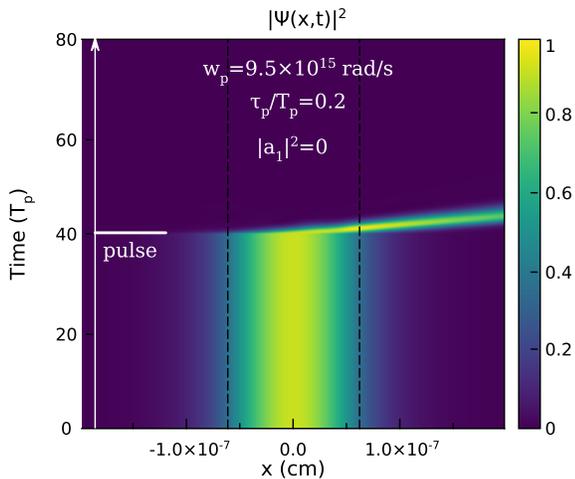}
\caption{The wave packet propagation in the one-dimensional QW with finite barriers of 0.2~eV. The ratio $\tau_{p}/T_{p}=0.2$. The dashed lines show the QW barriers. A magnitude of the wave function is normalized to unity for convenience.}
\label{figFinWP02}
\end{figure}

The wave packets for the QW with a single bound state excited by short unipolar pulses of the same amplitude $E_{0}=3\times 10^{8}$~V/cm and of different durations are shown in Figs.~\ref{figFinWP1},~\ref{figFinWP06}, and~\ref{figFinWP02}. From Figs.~\ref{figFinWP1} we see that for the excitation by a pulse of the duration $\tau_{p}=T_{p}$ there is no notable change of the bound state. Thus, its population remains close to unity. For the excitation by a pulse with $\tau_{p}=0.2\;T_{p}$ a complete ionization of the bound state takes place. It is clearly seen that the wave packet leaves the QW and goes outwards. As a result, the bound state becomes unpopulated. For the interim case $\tau_{p}=0.6\; T_{p}$, one can see that the ground state is fairly kept in the QW, though the oscillations representing harmonics of excited states appear. Thus, a partial excitation takes place, which is clearly indicated by the winding tails of the wave function after the pulse.

\section{Excitation of 3D nanostructures}

The one-dimensional consideration in the previous sections is just the simplest model of real nanostructures. However, as we show in this section, obtained results for 1D case allow an easy generalization to more realistic 3D geometry. Thus, here we model the nanostructure as a three-dimensional rectangular QW with infinite walls, i.e. the potential $U(x,y,z)=0,$ if $|x| \le l_1/2, \ \ |y| \le l_2/2, \ \ |z| \le l_3/2$, and $U(x,y,z) = \infty $ elsewhere. $l_1$, $l_2$, $l_3$ are the widths of the well in $x$-, $y$- and $z$-dimensions, respectively.

Eigenfunctions of the 3D QW are simply found as the product of three one-dimensional wave functions~(\ref{wf}):
\begin{eqnarray}
\Psi_{n_1,n_2,n_3}(x,y,z) = \Psi_{n_1}(x) \cdot \Psi_{n_2}(y) \cdot \Psi_{n_3}(z).
\label{wf3D}
\end{eqnarray}
The energy levels of 3D rectangular QW are given as~\cite{Landau}:
\begin{equation}
 E_n=\frac{\hbar^2}{2m} \left[ \left(\frac{\pi n_1}{l_1}\right)^2 + \left(\frac{\pi n_2}{l_2}\right)^2 + \left(\frac{\pi n_3}{l_3}\right)^2 \right].   
\end{equation}

Eq.~(\ref{sudden_solution}) for the wave function after excitation by a subcycle pulse now readily appears to be:
\begin{eqnarray}
\nonumber
\Psi(x,y,z) &=& \Psi_{111}(x,y,z) \cdot \exp{\Big( \frac{i}{\hbar} qxS_{E,x} \Big)} \cdot \\
&& \exp{\Big( \frac{i}{\hbar} qyS_{E,y} \Big)} \cdot 
\exp{\Big( \frac{i}{\hbar} qzS_{E,z} \Big)},
\label{psi3D}
\end{eqnarray}
where $\Psi_{1,1,1}(x,y,z)$ is the ground state wave function (\ref{wf3D}) of the unperturbed Hamiltonian.
We also introduced three scalar components $S_{E,x}$, $S_{E,y}$ and $S_{E,z}$ of the vector electric pulse area, $\vec S_{E} = \int_{t=-\infty }^{+\infty} \vec E(t) dt$. It is important to emphasize here, that we did not impose any restrictions on the polarization of the excitation pulse. Therefore Eq.~(\ref{psi3D}) is equally valid for an arbitrary polarized pump pulse.
Since all spatial variables in Eq.~(\ref{psi3D}) can be separated, it allows us to apply Eqs.~\eqref{amp_odd}-\eqref{amp_even}, obtained earlier for 1D case. Indeed, if we expand the wave function over the one-dimensional eigenfunctions, the expansion coefficients turn out to be:
\begin{eqnarray}
\nonumber
a_{n_1,n_2,n_3} &=& \int \int \int \Psi(x,y,z) \Psi_{n_1, n_2,n_3}(x,y,z) \ dx \ dy \ dz  \\
\nonumber
&=&  \int \Psi_1(x) \exp{\Big( \frac{i}{\hbar} qxS_{E,x} \Big)} \Psi_{n_1}(x) \ dx \cdot \\
\nonumber
&\cdot& \int \Psi_1(y) \exp{\Big( \frac{i}{\hbar} qyS_{E,y} \Big)} \Psi_{n_2}(y) \ dy \cdot \\
\nonumber
&\cdot& \int \Psi_1(z) \exp{\Big( \frac{i}{\hbar} qzS_{E,z} \Big)} \Psi_{n_3}(z) \ dz \\
&=& a^{1D}_{n_1} \cdot a^{1D}_{n_2} \cdot a^{1D}_{n_3},
\label{an3D}
\end{eqnarray}
where superscript ``1'' refers to Eqs.~\eqref{amp_odd}-\eqref{amp_even} for a 1D rectangular QW.

Eq.~\eqref{an3D} allows one to directly extend the parameter dependencies examined in the previous sections to the 3D case. In particular, Figs.~\ref{fig1}-\ref{fig2} describe the dependence of the level populations on the pulse duration $\tau_p$ for the case of a linearly-polarized pump pulse with the polarization direction along $x$-, $y$- or $z$-axis. If the electric field has several nonzero spatial components, one has to simply take the product of the corresponding one-dimensional dependencies on pulse parameters in order to obtain the respective dependence for a 3D QW.


\section{Conclusions}  
On the basis of an approximate  and direct numerical solution of TDSE, using theory of sudden perturbations, we studied  the interaction of subcycle attosecond pulses with a one-dimensional and three-dimensional QW nanostructures with finite and infinite barriers. Our results revealed, that populations of quantum states in the nanostructure can be controlled by the electric pulse area, when the pulse duration is smaller, than the characteristic time $T_g$ associated with the energy of the ground state in QW. In this case, it is shown  that the population of the bound states in QW are determined by the ratio of the electric pulse area to its characteristic scale proportional to the QW width.  For the subcycle pulses having nonzero electric area system remains in the excited state after pulse passes. In contrast, the impact of a zero-area single-cycle pulse does not lead to the excitation of the QW. 

The possibility of selective control of populations in QW by subcycle pulses is demonstrated in spite of the fact that the interaction is non-resonant. This selectivity allows to create population inversion in the structure on the resonant transition, which can be controlled by the pulse duration. These findings clearly demonstrate a possibility of the effective control of electron dynamics in QWs using half-cycle unipolar attosecond pulses through varying electric pulse area, which is in turn fully determined by the pulse duration, amplitude and CEP. A possibility to create population inversion can be used to obtain lasing in the nanostructures and shows the applicability of subcycle pulses for pumping. 

Our results revealed that the unipolar half-cycle attosecond pulses with large electric area can be used for the efficient ultrafast control of bound state population dynamics in nanostructures. Besides, their applications to the ultrafast control of quantum systems in nuclear physics can be also considered~\cite{Naumenko,arkhipovQW2023e}. The model of a QW with a single bound state is used for the description of the deuteron. Thus, for such a system, the estimation of the characteristic scale of the electric area $S_{D}$ can be performed.

In spite of the fact that first studies of unipolar half-cycle pulses were performed in 1960-s their existence remained questionable for many years. However, the results presented above show the importance of such pulses. The electric pulse area is an important quantity, which has evident physical meaning in this case, as it determines the momentum transferred to an electric charge and hence the excitation and ionization dynamics, when the pulse duration is shorter than $T_g$.  Therefore, we believe that our results presented above can connect two developing areas of modern optics - "attonanophysics" \cite{ciappina2017attosecond,ciappina2023multiphoton} and “optics of unipolar and subcycle light" \cite{arkhipov2020unipolar,arkhipov2022half,arkhipov2023unipolar,diachkova2023light}, which previously developed independently.

\section*{Acknowledgments}
This work was supported by Russian Science Foundation (project No. 21-72-10028). The calculations were performed using the facilities of the Resource Center ``Computational center of SPbU''.

\bibliography{UPlibrary}

\end{document}